\documentstyle[11pt]{article}

\def\ul{\underline}
     
\begin{document}

\begin{flushright}
BRX TH--392
\end{flushright}

\vspace{-.3in}

\begin{center}
{\large\bf The Quantum Field Theory of Physics 
and of Mathematics}\footnote{Invited talk at 
``Foundations of Quantum Field Theory," Boston 
University, Boston, MA, March 1--3, 1996.}

Howard J. Schnitzer\footnote{Supported in part 
by the DOE under grant DE-FG02-92ER40706}\\
Department of Physics\\
Brandeis University\\
Waltham, MA 02254
\end{center}

One recognizes that there has been, and continues 
to be, a great deal of common ground between 
statistical mechanics and quantum field theory 
(QFT).  Many of the effects and methods of 
statistical physics find parallels in QFT, 
particularly in the application of the latter 
to particle physics.  One encounters spontaneous 
symmetry breaking, renormalization group, solitons, 
effective field theories, fractional charge, and 
many other shared phenomena.

Professor Fisher \cite{Fisher} has given us a 
wonderful overview of the discovery and role of 
the renormalization group (RG) in statistical 
physics.  He also touched on some of the 
similarities and differences in the foundations 
of the RG in condensed matter and high energy 
physics, which was amplified in the discussion.  
In the latter subject, in addition to the 
formulation requiring cutoff-independence, we 
have the very fruitful Callan--Symanzik equations.  
That is, in the process of renormalizing the 
divergences of QFT, arbitrary, finite mass-scales 
appear in the renormalized amplitudes.  The 
Callan--Symanzik equations are the consequence of 
the requirement that the renormalized amplitudes 
in fact be independent of these arbitrary masses.  
This point of view is particularly useful in 
particle physics, although it does make its 
appearance in condensed matter physics as well.

The very beautiful subject of conformal field 
theory spans all three topics we are considering: 
critical phenomena, quantum field theory, and 
mathematics.  The relationship between conformal 
field theory and two-dimensional critical phenomena 
has become particularly fruitful in recent years.  
Conformal field theory in its own right, as well as 
being an aspect of string theory, has also played an 
important role in the recent vigorous interchanges 
between physics and mathematics.  A wide variety of 
mathematical techniques have become commonplace tools 
to physicists working in these areas.  Similarly, 
questions raised by conformal field theorists, 
topological field theorists, and string theorists have 
presented many new directions and opportunities for 
mathematics, particularly in geometry and topology.  
The relationship has been rich and symbiotic.

By contrast, it is my view that the more traditional 
mathematical physics of constructive and axiomatic 
field theory, and related endeavors, have not come 
any closer to the physicists' view of QFT.  In fact, 
with the increased appreciation and understanding of 
effective field theories, the gap between the two 
communities might even be widening, if I understand 
correctly.  It is this issue that I wish to address.

It should be clear that the quantum field theory of
mathematics is very different from that of physics.  
In fact, I have long had the opinion that these may 
not even be the same theories at all.  That is, there 
are (at least) two classes of quantum field
theories, which for historical reasons go by the same 
name, despite being very different.  The quantum 
field theory considered by the mathematics community 
is built on an axiomatic structure, and requires that 
the infinite volume system be consistent at \ul{all}
distance scales, infrared as well as
ultraviolet.  By contrast what physicists mean by 
a field theory
is, in contemporary language, an effective field  
theory, which is
applicable within a well-defined domain of validity, 
usually below some energy scale. 
Consistency is not required at short-distances, {\it
i.e.}, at energies above a specified energy scale.

Does any four dimensional QFT exist, in the 
mathematical sense?  The status of this topic was 
reviewed by Jaffe \cite{Jaffe}. As yet, no
four-dimensional QFT has been demonstrated to meet 
all the necessary requirements, although it is 
believed that pure Yang--Mills theory will eventually 
attain the status of a consistent theory.  What about 
the standard model?  There are two possibilities.  
Either it is just a matter of time for the necessary 
ingredients to be
assembled to produce a mathematically consistent 
four-dimensional
QFT describing the standard model, or no such theory 
exists.  Suppose a given candidate field theory of 
interest is shown in fact
not to be consistent.  One possible response is to 
embed the candidate theory
in a larger system with more degrees of freedom, 
{\it i.e.}, additional fields, and reexamine the 
consistency of the enlarged system.  The
hope is that eventually this sequence stops, and 
there is a consistent QFT.  However to repeat, the 
logical possibility exists
that this procedure does not lead to a consistent 
four-dimensional local QFT.

If no such consistent local QFT exists this 
would not have grave consequences for physical 
theory as we know it. From the physicists' point of 
view, the nesting of a sequence of theories is
a familiar strategy carrying a physical description 
to higher energies.
Embedding QED in the standard model, and then
into possible grand unified QFT, extends the 
description of fundamental interactions to at least 
100 Gev, and hopefully to still higher energies.  
However, the last \ul{field theory} in this sequence 
may not be consistent at
the shortest distance scales in the mathematical 
sense.  In any case, eventually in this
regression to shorter-distances, quantum effects of 
gravity are encountered.  The question of consistency 
then must change.  Then it becomes
plausible that to proceed to even shorter distances, 
something other than local QFT or \ul{local} quantum 
gravity will be required.  String theory provides 
one possibility for an extension to distances where 
quantum gravity is relevant.
The standard model is then just an effective 
low-energy representation of string theory.
It has even been speculated, on the basis of string
strong-weak duality, that string theory itself may 
only be an effective theory of some other theory  
(for example the as yet unknown M-theory)  \cite{Witten}.  
Therefore, the physicist does not (need to) care if 
there is any completely consistent local QFT, valid 
at the shortest distances.  The appropriate theories 
to be considered are effective theories; usually, 
but not always effective \ul{field} theories.

Jackiw \cite{Jackiw} argues elegantly and 
persuasively for  physical information carried by 
certain infinities of QFT; making their presence known 
in anomalies and spontaneous symmetry breaking.  
In QFT, the ultraviolet divergences should not just 
be regarded as awkward flaws, but rather a feature 
which can lead to physical consequences.  However it 
should not be concluded from his analysis that a 
theory without ultraviolet infinities
cannot describe the physical phenomena appropriate 
to anomalies and spontaneous symmetry breaking.  Such 
phenomena can be accommodated without ultraviolet 
divergences; string theory again
providing one example, although the language to 
describe the phenomena will differ.  Though 
ultraviolet divergences are an essential feature 
of local QFT, or effective field theory,
they are not necessary for a description of the 
physics.  Jackiw asks whether the string theory 
program has illuminated any physical questions. 
I should like to respond briefly in the affirmative.  
String theory has provided us with finite, well 
defined examples of quantum gravity; the only class 
of such theories presently known, with an essential 
aspect being non-locality.  It had long been 
conjectured that one could not construct a non-local 
theory which was Lorentz invariant, positive definite, 
causal and unitary.  One wondered whether there was 
any finite quantum gravity.  Certainly string theory 
sets these earlier prejudices aside and allows us to 
consider a much broader class of theories in 
confronting the physical world.  One should acknowledge 
that these are issues that have long been on the 
physicists' agenda, and are not solely of mathematical 
interest.  In this evolution of our understanding one 
still retains the basic assumptions of quantum 
mechanics, even though locality is no longer a sacred 
physical principle.

Mathematicians speak of field theories such as quantum
electrodynamics (QED) as heuristic field theories, or 
even as models, since no proof of mathematical 
consistency, at all
distances, exists in the sense mentioned earlier.  
I feel that this description of QED is pejorative, 
albeit unintended.  In fact
QED is the most precise physical theory ever 
constructed, with well-defined calculational rules 
for a very broad range of physical phenomena, and 
extraordinary experimental verification.
There is even a plausible explanation of why the 
fine-structure constant is small, based on a 
renormalization group extrapolation in grand unified 
theories. Of course, we understand that QED is an 
effective field theory, but is a well-defined theory 
in the sense of physical science. We know that to extend 
its domain of validity one
may embed it in the so-called standard model 
(itself an effective
field theory), the electro-weak sector of which 
has been tested over an enormous energy range 
(ev to 100 Gev; 11
decades), although not with the precision of QED.  
Thus both QED and the standard model are full-fledged 
theories of physical phenomena in every sense of 
the word!

The investigation of the mathematical consistency 
of four-dimensional local QFT is an interesting question 
in its own right.  No matter what the outcome, we will 
gain important insights into the structure of QFT.  
However, the answers to such questions are not likely 
to change the way we do particle physics.  Then how 
can mathematical physics make contact with issues of 
concern to particle physics?  What are the right 
questions?  Some suggestions immediately come to mind.
What is a mathematically
consistent \ul{effective} field theory? Is this even a 
well-posed problem?  If so, what restrictions does 
it place on effective
theories?  Can any candidates be discarded?  
To begin with, one should not expect that effective 
field theories are local field theories, as they 
involve infinite polynomials in fields and their 
derivatives.  Nor do they have to be consistent at 
the shortest distances.  An approach to some of these
issues has been made by Gomis and Weinberg 
\cite{Gomis} ``Are Non-renormalizable Gauge Theories 
Renormalizable in the Modern
Sense?"  They require that the infinities from
loop graphs be constrained by the symmetries of the 
bare action, such that there is a counterterm available 
to absorb every infinity.  This is a necessary 
requirement for a theory to make
sense perturbatively.  Their criteria are 
automatically satisfied
if the bare action arises from global, linearly 
realized symmetries.  However, this becomes a non-trivial 
requirement if either there are non-linearly realized 
symmetries or gauge  symmetries in the bare action.  
In constructive field theory one encounters a cutoff 
version of the local field theory being studied at 
intermediate stages of the analysis.  These can be 
regarded as effective field theories, but to be relevant 
they must be Lorentz invariant.  However, one needs to 
consider a wider class of effective theories than 
presently considered by constructive field theorists 
if the work is to have an impact on the concerns of 
particle physicists. In any case, there is certainly 
a great deal more to do in making effective field 
theories more precise.

In summary, I have argued that the QFT of 
mathematicians and of physicists are quite different, 
although both go by the name of QFT.  To bridge the 
gap, one recognizes that there are many
important mathematical problems posed by effective 
field theories, but these have not received the 
attention they deserve. Further, the existence of 
consistent string theories challenges the idea that 
locality is essential in the description of particle 
physics at the shortest distances.

\end{document}